\pdfoutput=1

\documentclass[%
 preprint,
 amsmath,amssymb,
]{revtex4-1}

\usepackage{graphicx}
\usepackage{dcolumn}
\usepackage{bm}
\usepackage[T1]{fontenc}      
\usepackage[greek,frenchb,english]{babel}

\newcommand{\YBCO}{$\mbox{YBa}_2$$\mbox{Cu}_3$$\mbox{O}_7$$\mbox{ }$}

\begin{document}

\title{Inductance measurement of YBCO strip-lines made by ion irradiation}

\author{T. Wolf}
\affiliation{Laboratoire de Physique et d'\'Etude des Mat\'eriaux - UMR 8213 ESPCI-UPMC-CNRS,
10, rue Vauquelin - 75231 Paris cedex 05, France}

\author{N. Bergeal}
\affiliation{Laboratoire de Physique et d'\'Etude des Mat\'eriaux - UMR 8213 ESPCI-UPMC-CNRS,
10, rue Vauquelin - 75231 Paris cedex 05, France}

\author{C. Ulysse}
\affiliation{Laboratoire de Photonique et Nanostructures - UPR 20 CNRS, route de Nozay, 91460 Marcoussis, France}

\author{G. Faini}
\affiliation{Laboratoire de Photonique et Nanostructures - UPR 20 CNRS, route de Nozay, 91460 Marcoussis, France}

\author{P. Febvre}
\affiliation{LAboratoire Hyperfr\'equences et Caract\'erisation - UMR 5130 CNRS, Universit\'e de Savoie, 73376 Le Bourget du Lac cedex, France}

\author{ J. Lesueur}
\email{jerome.lesueur@espci.fr}
\affiliation{Laboratoire de Physique et d'\'Etude des Mat\'eriaux - UMR 8213 ESPCI-UPMC-CNRS,
10, rue Vauquelin - 75231 Paris cedex 05, France}

\begin{abstract}
We have investigated the electrodynamic properties of High-Tc strip-lines
made by ion irradiation, in order to evaluate the potentialities of
such a technology for RSFQ superconductor digital electronic. SQUID
loops of different length and width have been fabricated by ion bombardment
of 70 nm thick films through e-beam lithographied shadow masks, and
measured at different temperatures. The voltage modulations have been
recorded by direct injection of a control current in the SQUIDs arms.
The corresponding line inductances have been measured and compared
with 3D simulations. A quantitative agreement has been obtained leading
to typical values of 0.4 pH/\textgreek{m}m without ground plane.
\end{abstract}
\maketitle

Superconductor digital electronics and its implementation of the Rapid
Single Flux Quantum (RSFQ) logic are currently the subject of intensive
research due to its exciting properties. It is assessed to be the
most advanced alternative technology to silicon-based systems in order
to reach the 100+ GHz operating frequency \cite{STA:2005}; it has
an extremely small energy consumption and its viability has been proven
through the successful development of highly complex low-Tc electronics
(for a review see \cite{Silver:2003,Fujimaki:2008}). For specific
applications like ADCs (Analog-to-Digital Converters), there is a
need to develop High-Temperature RSFQ devices, which would operate
beyond 400 GHz in the 30-80 K temperature range \cite{SCENET:2005,Tanabe:2008}.
While several competing technologies are currently been developped
for this goal, there has been recent interest in the ion-irradiated
Josephson Junction technology\cite{Bergeal:2005p1526,Bergeal:2006p1511,Bergeal:2007p1510}
which allows to design rather complex structures suitable for high
speed electronics \cite{Cybart:2009p6083}. In that context, mastering
the loop and line inductances in the circuits appears to be a key point.
In this paper, we present measurements of the inductance of superconducting
lines patterned using our all-planar ion-irradiation process\cite{Bergeal:2005p1526,Bergeal:2006p1511,Bergeal:2007p1510},
by direct current injection in SQUIDs arms. We compare the obtained
values with numerical simulation of the kinetic and magnetic components.
Given the good agreement observed, we then discuss the potential of
this technology for RSFQ and high-frequency devices.

\begin{figure}
\includegraphics[width=10cm]{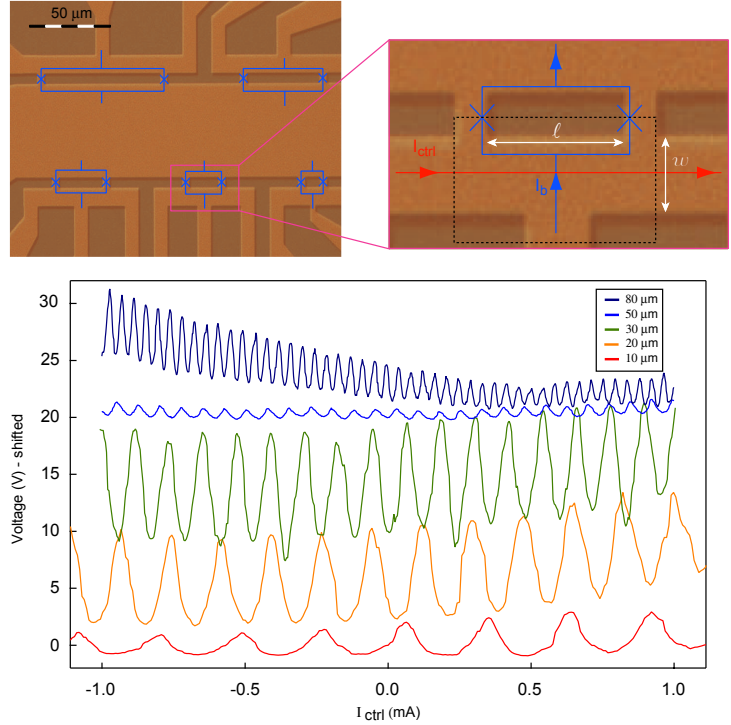}\caption{\label{fig:Figure 1} (Color online) Upper left panel : Picture of a sample before
the fabrication of the junctions. Five SQUIDs sharing a commun ground
electrode are seen, with length $l$ ranging from 10 to 80 \textgreek{m}m.
The control arm has a width $\mathit{w}$. Upper right panel : close
up of the $\mathit{l}$ = 20 \textgreek{m}m SQUID. A scheme of the
measurement geometry showing the bias current $I_{b}$ and the control
one $I_{ctrl}$ is displayed. The junctions are symbolized by a cross.
The dashed line shows the actual geometry used to compute the inductance
using the 3D-MLSI code. Lower panel : Voltage modulation (shifted
for clarity) for 10 \textgreek{m}m-wide SQUIDs as a function of the
control current $I{}_{ctrl}$ measured at 4.2 K. From bottom to top,
SQUIDs length are respectively : 80, 50, 30, 20 and 10 \textgreek{m}m.}

\end{figure}

Starting from a commercial 70 nm thick \YBCO (YBCO) film ($T_{c}$
= 86 K) grown on sapphire covered by an in-situ 100 nm gold layer,
a three steps fabrication process is performed. Firstly, contact pads
are defined in the gold layer through optical photoresist patterning
followed by a 500 eV Ar Ion Beam Etching (IBE). Secondly, contact
lines and SQUIDs arms are patterned in a AZ5214 image-reversal photoresist
followed by 110 keV oxygen ions irradiation. A dose of $5\times10^{15}$
ions/cm$^{\text{2}}$ ensures that the surrounding matrix is deeply
insulating. Eventually, Josephson junctions are fabricated : 20 nm
slits in a 600 nm thick poly(methylmethacrylate) (PMMA) photoresist
are patterned using a LEICA EBPG 5000+ electronic beamwriter and irradiated
with 110 keV oxygen ions. We used a dose of $3\times10^{13}$ ions/cm$^{\text{2}}$
suitable to operate in the temperature range 40 to 65 K. Different
geometries for the SQUIDs arms carrying the injection current have
been studied : width $\mathit{w}$ of 4 \textgreek{m}m and 10 \textgreek{m}m,
and five different lengths $\mathit{l}$ for each width : 10, 20,
30, 50 and 80 \textgreek{m}m. \ref{fig:Figure 1} displays the layout
of a typical sample showing five different SQUIDs (width of 10 \textgreek{m}m
and lengths from 10 to 80 \textgreek{m}m) sharing a common ground
electrode, together with current injection lines.

Measurements were conducted in an Oxford Variable Temperature Insert
using four-probe method and two extra-leads on one electrode of SQUIDs
for current injection in the superconducting wire (Upper right panel
of \ref{fig:Figure 1}). After measuring the critical current as a
function of T, the SQUID was biased at its optimal point ( $1.1\times I_{c}$
) while the control current was swept. This results in oscillations
in the SQUID's voltage (Lower panel in \ref{fig:Figure 1}) with
a current periodicity that can be related to the inductance of the
superconducting line through the simple formula $ $$\delta I_{ctrl}=\Phi_{0}/L$
where $\Phi_{0}=\frac{h}{2e}\simeq2\times10^{-15}$ Wb is the magnetic
flux quantum and $L$ the inductance of the superconducting line\cite{HASEGAWA:1995p2294,Ilichev:1996p6080,Fuke:1996p2296,Terai:1997p2300,Johansson:2009p2292}.

\begin{figure}
\includegraphics[bb=0bp 0bp 612bp 500bp,clip,width=10cm]{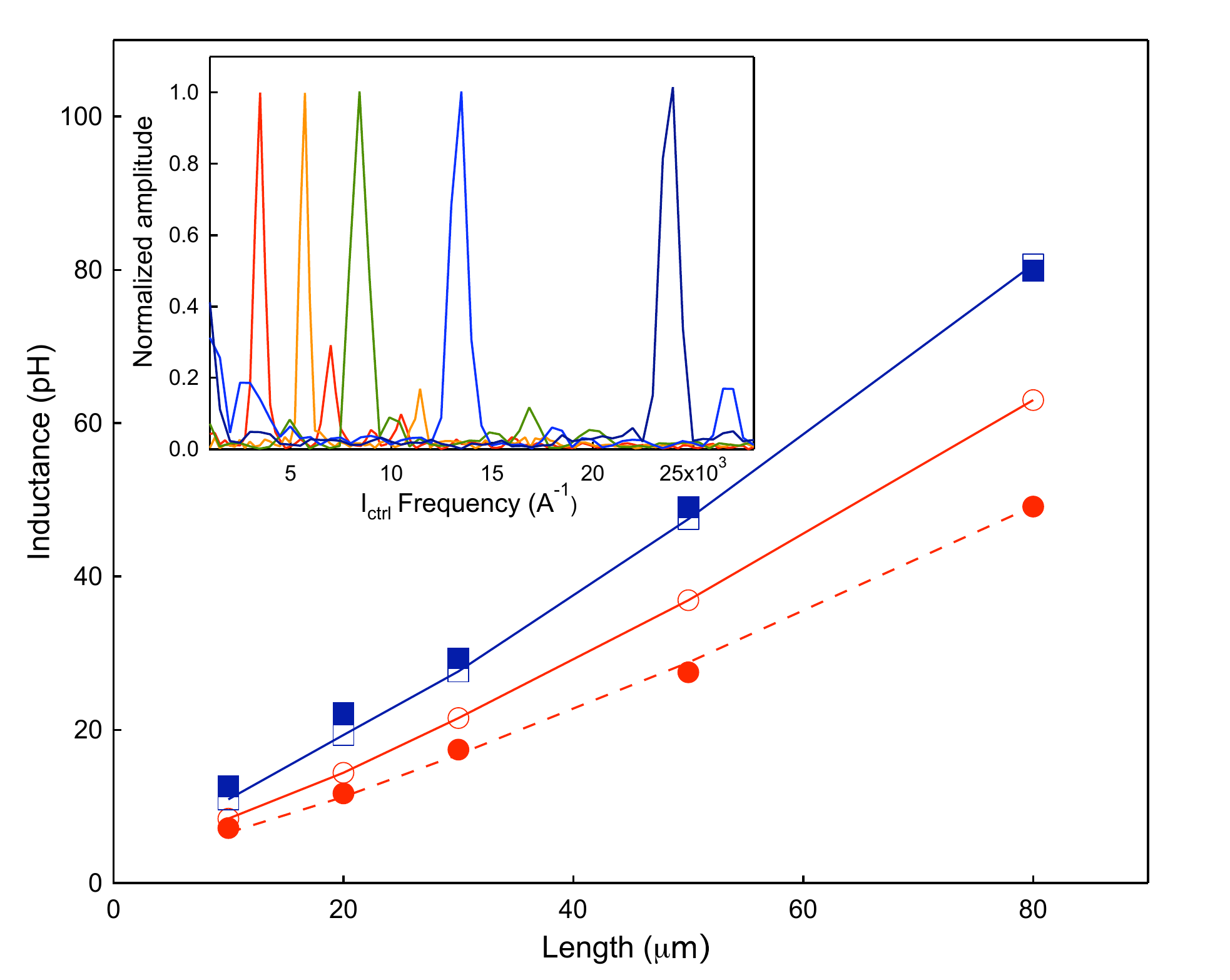}
\caption{\label{fig: Figure 2} (Color online) Inductance as a function of the nominal length
$\mathit{l}$ for two different widths ($\mathit{w}$ = 4 \textgreek{m}m
blue squares, and $w$ = 10 \textgreek{m}m red circles), measured
at 4.2 K for a $4.5\times10^{13}$ ions/cm$^{\text{2}}$ irradiated
sample. Open symbols linked by a solid line are the corresponding
values calculated with the 3D-MLSI code using $\lambda(0)$ = 0.135
\textgreek{m}m. The dashed line is the calculation including a correcting
parameter (see text). The inset displays the Fourier transform of
the SQUID modulations shown in Figure 1 for different lengths (from
left to right : 10, 20, 30, 50 and 80 \textgreek{m}m).}
\end{figure}

A first set of data was extracted from SQUIDs with highly irradiated
junctions measured at 4.2 K in order to extract the zero temperature
London penetration depth $\lambda(0)$. The injection current was
swept from -1 mA to 1 mA giving 20 to 50 oscillations of the SQUID
voltage which were enough for reliably Fourier transform the data
as displayed in the inset of \ref{fig: Figure 2}. The corresponding
inductances (solid symbols) are displayed in \ref{fig: Figure 2}
for the two widths as a function of the SQUID arm nominal length.
As expected, the total inductance increases slightly more rapidly
than linearly, since it is the sum of the kinetic contribution $L_{k}=\mu_{0}\lambda^{2}l/wt$
\cite{Duzer:1998p6846} and the geometric one $L_{m}=0.2l\left[1/2+\log(2l/(w+t))\right]$
\cite{Grover:1946p6847}, where $\lambda$ is the London penetration
depth and $t$ the thickness of the film. A more accurate estimate
of $L$ has been computed with the 3D-MLSI software \cite{Khapaev:2004p6848}
on the actual geometry of the SQUIDs (dashed line of figure \ref{fig:Figure 1}).
The results are given as open symbols in \ref{fig: Figure 2}. The
agreement for the 4 \textgreek{m}m wide superconducting lines is very
good given the zero temperature $\lambda(0)=0.135$ \textgreek{m}m,
which is exactly the film supplier's data, thus indicating that our
fabrication process preserves the superconducting properties of the
films. Concerning the 10 \textgreek{m}m wide superconducting lines,
the numerical simulations needs to be renormalized by a factor of
$G=0.78$ to yield to correct results (dashed line in \ref{fig: Figure 2}).
This numerical factor accounts for all the measured samples with this
geometry. We do not have a clear explanation for this 20\% discrepancy.
However we should stress that most of the reported data in the literature
using this DC SQUID method, do not provide quantitative agreements
between measured and calculated inductances with such a precision.

\begin{figure}
\includegraphics[width=10cm]{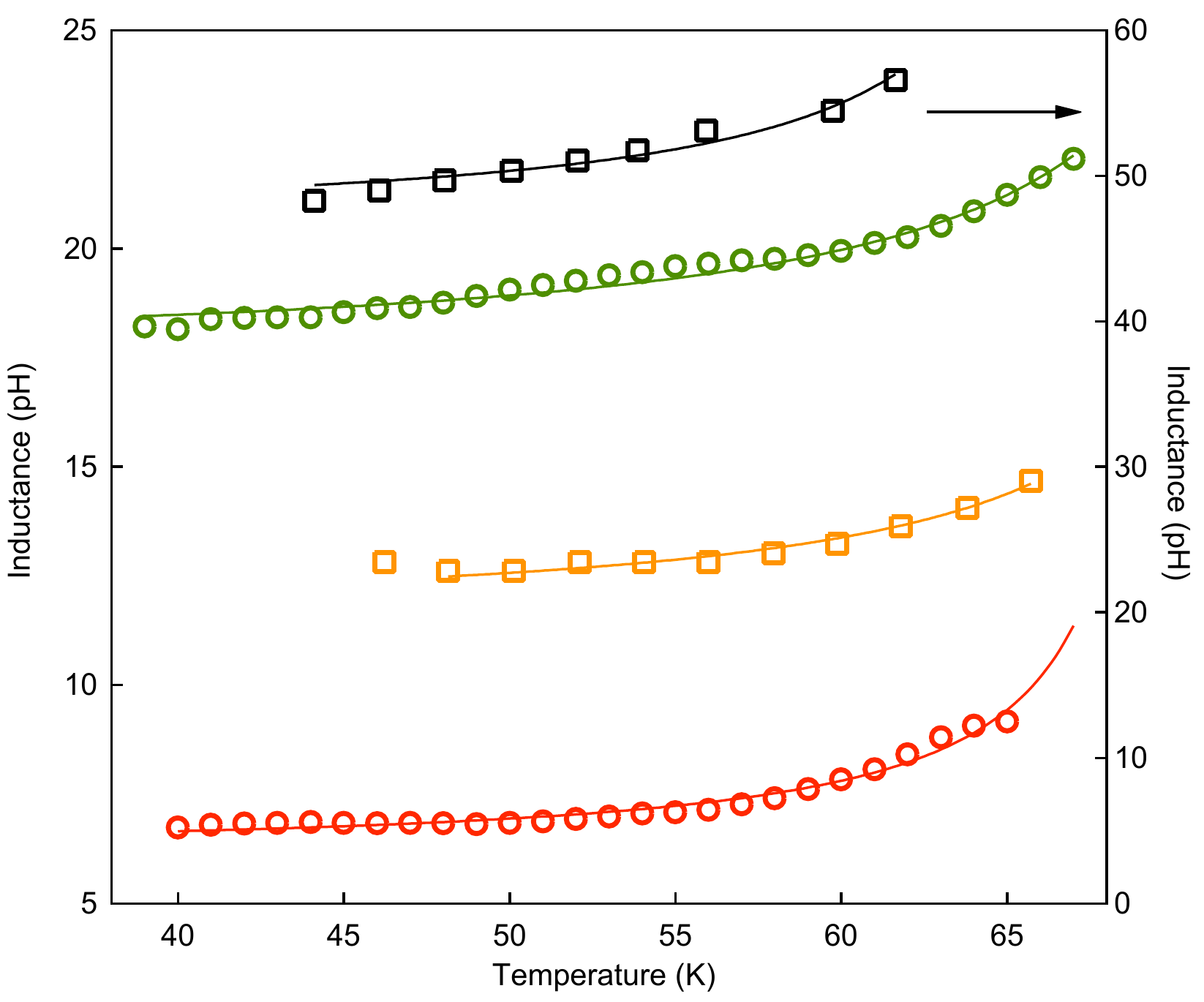}

\caption{\label{fig: Figure 3} (Color online) Inductances as a function of temperature measured
for a $3\times10^{13}$ ions/cm$^{\text{2}}$ irradiated sample of
width $\mathit{w}$ = 10 \textgreek{m}m, and length 10, 20, 30 and
80 \textgreek{m}m from bottom to top. The solid lines are the best
fit to the data with $\lambda(0)$ = 0.135 \textgreek{m}m and the
correction factor $G$ = 0.78 (see text), with respective $T_{c}$
of 70.5 K, 75.5 K, 75.5 K and 69 K.}

\end{figure}

The second set of SQUIDs were operated in the temperature range of
interest 40-65 K and allowed us to extract the temperature dependence
of the London penetration depth. The injection current was swept up
to 10 mA to give the precision needed to measure the slight variation
in inductance with temperature. \ref{fig: Figure 3} summarizes our
results together with fits of the temperature dependence of the inductance
controlled by the penetration depth. Here we have used the Gorter-Casimir
two fluids models with an exponent \textgreek{a} = 2: $\lambda(T)=\lambda(0)\frac{1}{\sqrt{1-\left(\frac{T}{T_{c}}\right)^{\alpha}}}$
\cite{Ilichev:1996p6080,Terai:1997p2300,Johansson:2009p2292}. Given
the values $\lambda(0)$ = 0.135 \textgreek{m}m and $G$ = 0.78 measured
previously for the 10 \textgreek{m}m wide sample, we found a good
agreement providing that $T_{c}$ is adjusted within 5 K or so (see
caption of \ref{fig: Figure 3}). This uncertaincy has two origins
related to the measurement method itself. Firstly, the Josephson regime
of the irradiated junctions is limited in temperature\cite{Bergeal:2005p1526}.
Therefore, our measurements are restricted in a temperature range
(40-65 K in this case) well below the bulk $T_{c}$, in a region where
$\lambda(T)$ does not strongly vary. Secondly, as the sensitivity
of the measurement decreases when we approach this coupling temperature,
one has to use high current density exceeding 10$^{\text{6}}$ A/cm$^{\text{2}}$
, that is the critical one in our samples at 77 K. The actual $T_{c}$
in the wire might therefore be lowered.

The overall measurements of inductances give us a value of 0.6 pH/\textgreek{m}m
for a 10 \textgreek{m}m wide line and 0.9 pH/\textgreek{m}m for a
4 \textgreek{m}m wide line. These numbers compare favorably with the
best one reported in the literature \cite{Terai:1997p2300}. They
do not change if the strip-line is embedded in a coplanar wave guide
geometry with a typical gap of 4 \textgreek{m}m with respect to the
ground plane. However, a factor of three improvement is expected when
a ground plane is added \cite{Forrester:1994p2299,Terai:1997p2300}.
Such a desirable situation could be realized by adapting the ion irradiation
process to trilayer films \cite{Bergeal:2007p1510,Bergeal:2008}.

Ion-irradiated junctions have typical critical current values in the
range 100 \textgreek{m}A - 1 mA which leads to a screening parameter
$2LI_{c}/\Phi_{0}$ ranging between 0.6 and 6 for a 10 \textgreek{m}m
long and wide strip, or 0.36 and 3.6 for a 4 \textgreek{m}m one. Since
RSFQ logic requires screening parameters around 1 for transmission
lines and around 3 for storage cell \cite{Gueret:1980,Semenov:1991},
the above mentioned inductances values fit well. Therefore, the irradiated
junction technology seems suitable for RSFQ circuits applications.

In conclusion, we have measured the inductance of YBCO strip lines
made by ion irradiation using direct current injection in SQUIDs arms.
The measured values agree well with the calculated one using 3D simulation
of our circuits. Such a line inductance combined with the typical
critical currents of ion irradiated Josephson Junctions opens the
route towards the realization of RSFQ logic circuits using this new
technology.\\

This work has been supported by the Region Ile-de-France in the framework
of C\textquoteright{}Nano IdF and by the Délégation Générale de l'Armement
(DGA) through a doctoral grant. C\textquoteright{}Nano IdF is the
nanoscience competence center of Paris Region, supported by CNRS,
CEA, MESR and Region Ile-de-France. Authors would like to thank Y.
Legall at INESS Strasbourg for the ion irradiations.



\begin{thebibliography}{21}%
\makeatletter
\providecommand \@ifxundefined [1]{%
 \@ifx{#1\undefined}
}%
\providecommand \@ifnum [1]{%
 \ifnum #1\expandafter \@firstoftwo
 \else \expandafter \@secondoftwo
 \fi
}%
\providecommand \@ifx [1]{%
 \ifx #1\expandafter \@firstoftwo
 \else \expandafter \@secondoftwo
 \fi
}%
\providecommand \natexlab [1]{#1}%
\providecommand \enquote  [1]{``#1''}%
\providecommand \bibnamefont  [1]{#1}%
\providecommand \bibfnamefont [1]{#1}%
\providecommand \citenamefont [1]{#1}%
\providecommand \href@noop [0]{\@secondoftwo}%
\providecommand \href [0]{\begingroup \@sanitize@url \@href}%
\providecommand \@href[1]{\@@startlink{#1}\@@href}%
\providecommand \@@href[1]{\endgroup#1\@@endlink}%
\providecommand \@sanitize@url [0]{\catcode `\\12\catcode `\$12\catcode
  `\&12\catcode `\#12\catcode `\^12\catcode `\_12\catcode `\%12\relax}%
\providecommand \@@startlink[1]{}%
\providecommand \@@endlink[0]{}%
\providecommand \url  [0]{\begingroup\@sanitize@url \@url }%
\providecommand \@url [1]{\endgroup\@href {#1}{\urlprefix }}%
\providecommand \urlprefix  [0]{URL }%
\providecommand \Eprint [0]{\href }%
\@ifxundefined \urlstyle {%
  \providecommand \doi  [0]{\begingroup \@sanitize@url \@doi}%
  \providecommand \@doi [1]{\endgroup \@@startlink {\doibase
  #1}doi:\discretionary {}{}{}#1\@@endlink }%
}{%
  \providecommand \doi  [0]{doi:\discretionary{}{}{}\begingroup
  \urlstyle{rm}\Url }%
}%
\providecommand \doibase [0]{http://dx.doi.org/}%
\providecommand \Doi [0]{\begingroup \@sanitize@url \@Doi }%
\providecommand \@Doi  [1]{\endgroup\@@startlink{\doibase#1}\@@Doi}%
\providecommand \@@Doi [1]{#1\@@endlink}%
\providecommand \selectlanguage [0]{\@gobble}%
\providecommand \bibinfo  [0]{\@secondoftwo}%
\providecommand \bibfield  [0]{\@secondoftwo}%
\providecommand \translation [1]{[#1]}%
\providecommand \BibitemOpen [0]{}%
\providecommand \bibitemStop [0]{}%
\providecommand \bibitemNoStop [0]{.\EOS\space}%
\providecommand \EOS [0]{\spacefactor3000\relax}%
\providecommand \BibitemShut  [1]{\csname bibitem#1\endcsname}%
\bibitem [{\citenamefont {Bedard}\ \emph {et~al.}(2005)\citenamefont {Bedard},
  \citenamefont {Welker}, \citenamefont {Cotter}, \citenamefont {Escavage},\
  and\ \citenamefont {Pinkston}}]{STA:2005}%
  \BibitemOpen
  \bibfield  {author} {\bibinfo {author} {\bibfnamefont {F.}~\bibnamefont
  {Bedard}}, \bibinfo {author} {\bibfnamefont {N.~K.}\ \bibnamefont {Welker}},
  \bibinfo {author} {\bibfnamefont {G.~R.}\ \bibnamefont {Cotter}}, \bibinfo
  {author} {\bibfnamefont {M.~A.}\ \bibnamefont {Escavage}}, \ and\ \bibinfo
  {author} {\bibfnamefont {J.~T.}\ \bibnamefont {Pinkston}},\ }\href@noop {}
  {\bibfield  {journal} {\bibinfo  {journal} {National Security Agency Office
  of Corporate Assessments}} (\bibinfo {year} {2005})}\BibitemShut {NoStop}%
\bibitem [{\citenamefont {Silver}\ \emph {et~al.}(2003)\citenamefont {Silver},
  \citenamefont {Kleinsasser}, \citenamefont {Kerber}, \citenamefont {Herr},
  \citenamefont {Dorojevets}, \citenamefont {Bunyk},\ and\ \citenamefont
  {Abelson}}]{Silver:2003}%
  \BibitemOpen
  \bibfield  {author} {\bibinfo {author} {\bibfnamefont {A.}~\bibnamefont
  {Silver}}, \bibinfo {author} {\bibfnamefont {A.}~\bibnamefont {Kleinsasser}},
  \bibinfo {author} {\bibfnamefont {G.}~\bibnamefont {Kerber}}, \bibinfo
  {author} {\bibfnamefont {Q.}~\bibnamefont {Herr}}, \bibinfo {author}
  {\bibfnamefont {M.}~\bibnamefont {Dorojevets}}, \bibinfo {author}
  {\bibfnamefont {P.}~\bibnamefont {Bunyk}}, \ and\ \bibinfo {author}
  {\bibfnamefont {L.}~\bibnamefont {Abelson}},\ }\href@noop {} {\bibfield
  {journal} {\bibinfo  {journal} {Superconductor Science and Technology},\
  }\textbf {\bibinfo {volume} {16}},\ \bibinfo {pages} {1368} (\bibinfo {year}
  {2003})}\BibitemShut {NoStop}%
\bibitem [{\citenamefont {Fujimaki}\ \emph {et~al.}(2008)\citenamefont
  {Fujimaki}, \citenamefont {Tanaka}, \citenamefont {Yamada}, \citenamefont
  {Yamanashi}, \citenamefont {Park},\ and\ \citenamefont
  {Yoshikawa}}]{Fujimaki:2008}%
  \BibitemOpen
  \bibfield  {author} {\bibinfo {author} {\bibfnamefont {A.}~\bibnamefont
  {Fujimaki}}, \bibinfo {author} {\bibfnamefont {M.}~\bibnamefont {Tanaka}},
  \bibinfo {author} {\bibfnamefont {T.}~\bibnamefont {Yamada}}, \bibinfo
  {author} {\bibfnamefont {Y.}~\bibnamefont {Yamanashi}}, \bibinfo {author}
  {\bibfnamefont {H.}~\bibnamefont {Park}}, \ and\ \bibinfo {author}
  {\bibfnamefont {N.}~\bibnamefont {Yoshikawa}},\ }\href@noop {} {\bibfield
  {journal} {\bibinfo  {journal} {IEICE Transactions on Electronics},\ }\textbf
  {\bibinfo {volume} {E91-C}},\ \bibinfo {pages} {342} (\bibinfo {year}
  {2008})}\BibitemShut {NoStop}%
\bibitem [{\citenamefont {ter Brake}\ \emph {et~al.}(2008)\citenamefont {ter
  Brake}, \citenamefont {Buchholz}, \citenamefont {Burnell}, \citenamefont
  {Claeson}, \citenamefont {Cr{\'e}te}, \citenamefont {Febvre}, \citenamefont
  {Gerritsma}, \citenamefont {Hilgenkamp}, \citenamefont {Humphreys},
  \citenamefont {Ivanov}, \citenamefont {Jutzi}, \citenamefont {Khabipov},
  \citenamefont {Mannhart}, \citenamefont {Meyer}, \citenamefont {Niemeyer},
  \citenamefont {Ravex}, \citenamefont {Rogalla}, \citenamefont {Russo},
  \citenamefont {Satchell}, \citenamefont {Siegel}, \citenamefont {T{\"o}pfer},
  \citenamefont {Uhlmann}, \citenamefont {Vill{\'e}gier}, \citenamefont
  {Wikborg}, \citenamefont {Winkler},\ and\ \citenamefont
  {Zorin}}]{SCENET:2005}%
  \BibitemOpen
  \bibfield  {author} {\bibinfo {author} {\bibfnamefont {H.}~\bibnamefont {ter
  Brake}}, \bibinfo {author} {\bibfnamefont {F.-I.}\ \bibnamefont {Buchholz}},
  \bibinfo {author} {\bibfnamefont {G.}~\bibnamefont {Burnell}}, \bibinfo
  {author} {\bibfnamefont {T.}~\bibnamefont {Claeson}}, \bibinfo {author}
  {\bibfnamefont {D.}~\bibnamefont {Cr{\'e}te}}, \bibinfo {author}
  {\bibfnamefont {P.}~\bibnamefont {Febvre}}, \bibinfo {author} {\bibfnamefont
  {G.}~\bibnamefont {Gerritsma}}, \bibinfo {author} {\bibfnamefont
  {H.}~\bibnamefont {Hilgenkamp}}, \bibinfo {author} {\bibfnamefont
  {R.}~\bibnamefont {Humphreys}}, \bibinfo {author} {\bibfnamefont
  {Z.}~\bibnamefont {Ivanov}}, \bibinfo {author} {\bibfnamefont
  {W.}~\bibnamefont {Jutzi}}, \bibinfo {author} {\bibfnamefont
  {M.}~\bibnamefont {Khabipov}}, \bibinfo {author} {\bibfnamefont
  {J.}~\bibnamefont {Mannhart}}, \bibinfo {author} {\bibfnamefont {H.-G.}\
  \bibnamefont {Meyer}}, \bibinfo {author} {\bibfnamefont {J.}~\bibnamefont
  {Niemeyer}}, \bibinfo {author} {\bibfnamefont {A.}~\bibnamefont {Ravex}},
  \bibinfo {author} {\bibfnamefont {H.}~\bibnamefont {Rogalla}}, \bibinfo
  {author} {\bibfnamefont {M.}~\bibnamefont {Russo}}, \bibinfo {author}
  {\bibfnamefont {J.}~\bibnamefont {Satchell}}, \bibinfo {author}
  {\bibfnamefont {M.}~\bibnamefont {Siegel}}, \bibinfo {author} {\bibfnamefont
  {H.}~\bibnamefont {T{\"o}pfer}}, \bibinfo {author} {\bibfnamefont
  {F.}~\bibnamefont {Uhlmann}}, \bibinfo {author} {\bibfnamefont {J.-C.}\
  \bibnamefont {Vill{\'e}gier}}, \bibinfo {author} {\bibfnamefont
  {E.}~\bibnamefont {Wikborg}}, \bibinfo {author} {\bibfnamefont
  {D.}~\bibnamefont {Winkler}}, \ and\ \bibinfo {author} {\bibfnamefont
  {A.}~\bibnamefont {Zorin}},\ }\href@noop {} {\bibfield  {journal} {\bibinfo
  {journal} {Physica C},\ }\textbf {\bibinfo {volume} {439}},\ \bibinfo {pages}
  {1} (\bibinfo {year} {2008})}\BibitemShut {NoStop}%
\bibitem [{\citenamefont {Tanabe}\ \emph {et~al.}(2008)\citenamefont {Tanabe},
  \citenamefont {Wakana}, \citenamefont {Tsubone}, \citenamefont {Tarutani},
  \citenamefont {Adachi}, \citenamefont {Ishimaru}, \citenamefont {Maruyama},
  \citenamefont {Hato}, \citenamefont {Yoshida},\ and\ \citenamefont
  {Suzuki}}]{Tanabe:2008}%
  \BibitemOpen
  \bibfield  {author} {\bibinfo {author} {\bibfnamefont {K.}~\bibnamefont
  {Tanabe}}, \bibinfo {author} {\bibfnamefont {H.}~\bibnamefont {Wakana}},
  \bibinfo {author} {\bibfnamefont {K.}~\bibnamefont {Tsubone}}, \bibinfo
  {author} {\bibfnamefont {Y.}~\bibnamefont {Tarutani}}, \bibinfo {author}
  {\bibfnamefont {S.}~\bibnamefont {Adachi}}, \bibinfo {author} {\bibfnamefont
  {Y.}~\bibnamefont {Ishimaru}}, \bibinfo {author} {\bibfnamefont
  {M.}~\bibnamefont {Maruyama}}, \bibinfo {author} {\bibfnamefont
  {T.}~\bibnamefont {Hato}}, \bibinfo {author} {\bibfnamefont {A.}~\bibnamefont
  {Yoshida}}, \ and\ \bibinfo {author} {\bibfnamefont {H.}~\bibnamefont
  {Suzuki}},\ }\href@noop {} {\bibfield  {journal} {\bibinfo  {journal} {IEICE
  Trans. Electron.},\ }\textbf {\bibinfo {volume} {E91-C}},\ \bibinfo {pages}
  {280} (\bibinfo {year} {2008})}\BibitemShut {NoStop}%
\bibitem [{\citenamefont {Bergeal}\ \emph {et~al.}(2005)\citenamefont
  {Bergeal}, \citenamefont {Grison}, \citenamefont {Lesueur}, \citenamefont
  {Faini}, \citenamefont {Aprili},\ and\ \citenamefont
  {Contour}}]{Bergeal:2005p1526}%
  \BibitemOpen
  \bibfield  {author} {\bibinfo {author} {\bibfnamefont {N.}~\bibnamefont
  {Bergeal}}, \bibinfo {author} {\bibfnamefont {X.}~\bibnamefont {Grison}},
  \bibinfo {author} {\bibfnamefont {J.}~\bibnamefont {Lesueur}}, \bibinfo
  {author} {\bibfnamefont {G.}~\bibnamefont {Faini}}, \bibinfo {author}
  {\bibfnamefont {M.}~\bibnamefont {Aprili}}, \ and\ \bibinfo {author}
  {\bibfnamefont {J.}~\bibnamefont {Contour}},\ }\href@noop {} {\bibfield
  {journal} {\bibinfo  {journal} {Applied Physics Letters},\ }\textbf {\bibinfo
  {volume} {87}},\ \bibinfo {pages} {102502} (\bibinfo {year}
  {2005})}\BibitemShut {NoStop}%
\bibitem [{\citenamefont {Bergeal}\ \emph {et~al.}(2006)\citenamefont
  {Bergeal}, \citenamefont {Lesueur}, \citenamefont {Faini}, \citenamefont
  {Aprili},\ and\ \citenamefont {Contour}}]{Bergeal:2006p1511}%
  \BibitemOpen
  \bibfield  {author} {\bibinfo {author} {\bibfnamefont {N.}~\bibnamefont
  {Bergeal}}, \bibinfo {author} {\bibfnamefont {J.}~\bibnamefont {Lesueur}},
  \bibinfo {author} {\bibfnamefont {G.}~\bibnamefont {Faini}}, \bibinfo
  {author} {\bibfnamefont {M.}~\bibnamefont {Aprili}}, \ and\ \bibinfo {author}
  {\bibfnamefont {J.~P.}\ \bibnamefont {Contour}},\ }\href@noop {} {\bibfield
  {journal} {\bibinfo  {journal} {Applied Physics Letters},\ }\textbf {\bibinfo
  {volume} {89}},\ \bibinfo {pages} {112515} (\bibinfo {year}
  {2006})}\BibitemShut {NoStop}%
\bibitem [{\citenamefont {Bergeal}\ \emph {et~al.}(2007)\citenamefont
  {Bergeal}, \citenamefont {Lesueur}, \citenamefont {Sirena}, \citenamefont
  {Faini}, \citenamefont {Aprili}, \citenamefont {Contour},\ and\ \citenamefont
  {Leridon}}]{Bergeal:2007p1510}%
  \BibitemOpen
  \bibfield  {author} {\bibinfo {author} {\bibfnamefont {N.}~\bibnamefont
  {Bergeal}}, \bibinfo {author} {\bibfnamefont {J.}~\bibnamefont {Lesueur}},
  \bibinfo {author} {\bibfnamefont {M.}~\bibnamefont {Sirena}}, \bibinfo
  {author} {\bibfnamefont {G.}~\bibnamefont {Faini}}, \bibinfo {author}
  {\bibfnamefont {M.}~\bibnamefont {Aprili}}, \bibinfo {author} {\bibfnamefont
  {J.~P.}\ \bibnamefont {Contour}}, \ and\ \bibinfo {author} {\bibfnamefont
  {B.}~\bibnamefont {Leridon}},\ }\href@noop {} {\bibfield  {journal} {\bibinfo
   {journal} {J Appl Phys},\ }\textbf {\bibinfo {volume} {102}},\ \bibinfo
  {pages} {083903} (\bibinfo {year} {2007})}\BibitemShut {NoStop}%
\bibitem [{\citenamefont {Cybart}\ \emph {et~al.}(2009)\citenamefont {Cybart},
  \citenamefont {Anton}, \citenamefont {Wu}, \citenamefont {Clarke},\ and\
  \citenamefont {Dynes}}]{Cybart:2009p6083}%
  \BibitemOpen
  \bibfield  {author} {\bibinfo {author} {\bibfnamefont {S.~A.}\ \bibnamefont
  {Cybart}}, \bibinfo {author} {\bibfnamefont {S.~M.}\ \bibnamefont {Anton}},
  \bibinfo {author} {\bibfnamefont {S.~M.}\ \bibnamefont {Wu}}, \bibinfo
  {author} {\bibfnamefont {J.}~\bibnamefont {Clarke}}, \ and\ \bibinfo {author}
  {\bibfnamefont {R.~C.}\ \bibnamefont {Dynes}},\ }\href@noop {} {\bibfield
  {journal} {\bibinfo  {journal} {Nano Lett},\ }\textbf {\bibinfo {volume}
  {9}},\ \bibinfo {pages} {3581} (\bibinfo {year} {2009})}\BibitemShut
  {NoStop}%
\bibitem [{\citenamefont {Hasegawa}\ \emph {et~al.}(1995)\citenamefont
  {Hasegawa}, \citenamefont {Tarutani}, \citenamefont {Fukazawa}, \citenamefont
  {Kabasawa},\ and\ \citenamefont {Takagi}}]{HASEGAWA:1995p2294}%
  \BibitemOpen
  \bibfield  {author} {\bibinfo {author} {\bibfnamefont {H.}~\bibnamefont
  {Hasegawa}}, \bibinfo {author} {\bibfnamefont {Y.}~\bibnamefont {Tarutani}},
  \bibinfo {author} {\bibfnamefont {T.}~\bibnamefont {Fukazawa}}, \bibinfo
  {author} {\bibfnamefont {U.}~\bibnamefont {Kabasawa}}, \ and\ \bibinfo
  {author} {\bibfnamefont {K.}~\bibnamefont {Takagi}},\ }\href@noop {}
  {\bibfield  {journal} {\bibinfo  {journal} {Applied Physics Letters},\
  }\textbf {\bibinfo {volume} {67}},\ \bibinfo {pages} {3177} (\bibinfo {year}
  {1995})}\BibitemShut {NoStop}%
\bibitem [{\citenamefont {Il'ichev}\ \emph {et~al.}(1996)\citenamefont
  {Il'ichev}, \citenamefont {D{\"o}rrer}, \citenamefont {Schmidl},\ and\
  \citenamefont {Zakosarenko}}]{Ilichev:1996p6080}%
  \BibitemOpen
  \bibfield  {author} {\bibinfo {author} {\bibfnamefont {E.}~\bibnamefont
  {Il'ichev}}, \bibinfo {author} {\bibfnamefont {L.}~\bibnamefont
  {D{\"o}rrer}}, \bibinfo {author} {\bibfnamefont {F.}~\bibnamefont {Schmidl}},
  \ and\ \bibinfo {author} {\bibfnamefont {V.}~\bibnamefont {Zakosarenko}},\
  }\href@noop {} {\bibfield  {journal} {\bibinfo  {journal} {Applied Physics
  {\ldots}}} (\bibinfo {year} {1996})}\BibitemShut {NoStop}%
\bibitem [{\citenamefont {Fuke}\ \emph {et~al.}(1996)\citenamefont {Fuke},
  \citenamefont {Saitoh}, \citenamefont {Utagawa},\ and\ \citenamefont
  {Enomoto}}]{Fuke:1996p2296}%
  \BibitemOpen
  \bibfield  {author} {\bibinfo {author} {\bibfnamefont {H.}~\bibnamefont
  {Fuke}}, \bibinfo {author} {\bibfnamefont {K.}~\bibnamefont {Saitoh}},
  \bibinfo {author} {\bibfnamefont {T.}~\bibnamefont {Utagawa}}, \ and\
  \bibinfo {author} {\bibfnamefont {Y.}~\bibnamefont {Enomoto}},\ }\href@noop
  {} {\bibfield  {journal} {\bibinfo  {journal} {Japanese Journal Of Applied
  Physics},\ }\textbf {\bibinfo {volume} {35}},\ \bibinfo {pages} {1582}
  (\bibinfo {year} {1996})}\BibitemShut {NoStop}%
\bibitem [{\citenamefont {Terai}\ \emph {et~al.}(1997)\citenamefont {Terai},
  \citenamefont {Hidaka}, \citenamefont {Satoh},\ and\ \citenamefont
  {Tahara}}]{Terai:1997p2300}%
  \BibitemOpen
  \bibfield  {author} {\bibinfo {author} {\bibfnamefont {H.}~\bibnamefont
  {Terai}}, \bibinfo {author} {\bibfnamefont {M.}~\bibnamefont {Hidaka}},
  \bibinfo {author} {\bibfnamefont {T.}~\bibnamefont {Satoh}}, \ and\ \bibinfo
  {author} {\bibfnamefont {S.}~\bibnamefont {Tahara}},\ }\href@noop {}
  {\bibfield  {journal} {\bibinfo  {journal} {Applied Physics Letters},\
  }\textbf {\bibinfo {volume} {70}},\ \bibinfo {pages} {2690} (\bibinfo {year}
  {1997})}\BibitemShut {NoStop}%
\bibitem [{\citenamefont {Johansson}\ \emph {et~al.}(2009)\citenamefont
  {Johansson}, \citenamefont {Cedergren},\ and\ \citenamefont
  {Bauch}}]{Johansson:2009p2292}%
  \BibitemOpen
  \bibfield  {author} {\bibinfo {author} {\bibfnamefont {J.}~\bibnamefont
  {Johansson}}, \bibinfo {author} {\bibfnamefont {K.}~\bibnamefont
  {Cedergren}}, \ and\ \bibinfo {author} {\bibfnamefont {T.}~\bibnamefont
  {Bauch}},\ }\href@noop {} {\bibfield  {journal} {\bibinfo  {journal}
  {Physical Review B}} (\bibinfo {year} {2009})}\BibitemShut {NoStop}%
\bibitem [{\citenamefont {Duzer}\ and\ \citenamefont
  {Turner}(1998)}]{Duzer:1998p6846}%
  \BibitemOpen
  \bibfield  {author} {\bibinfo {author} {\bibfnamefont {T.~V.}\ \bibnamefont
  {Duzer}}\ and\ \bibinfo {author} {\bibfnamefont {C.}~\bibnamefont {Turner}},\
  }\href@noop {} {\bibfield  {journal} {\bibinfo  {journal} {New York: Elsevier
  North Holland},\ }\textbf {\bibinfo {volume} {Principles of Superconductive
  Devices and Circuits}} (\bibinfo {year} {1998})}\BibitemShut {NoStop}%
\bibitem [{\citenamefont {Grover}(1946)}]{Grover:1946p6847}%
  \BibitemOpen
  \bibfield  {author} {\bibinfo {author} {\bibfnamefont {F.}~\bibnamefont
  {Grover}},\ }\href@noop {} {\bibfield  {journal} {\bibinfo  {journal} {New
  York: Dover Publications},\ }\textbf {\bibinfo {volume} {Inductance
  calculations - Working Formulas and Tables}} (\bibinfo {year}
  {1946})}\BibitemShut {NoStop}%
\bibitem [{\citenamefont {Khapaev}\ and\ \citenamefont
  {Goldobin}(2004)}]{Khapaev:2004p6848}%
  \BibitemOpen
  \bibfield  {author} {\bibinfo {author} {\bibfnamefont {M.}~\bibnamefont
  {Khapaev}}\ and\ \bibinfo {author} {\bibfnamefont {E.}~\bibnamefont
  {Goldobin}},\ }\href@noop {} {\bibfield  {journal} {\bibinfo  {journal}
  {http://www.cmc.msu.ru/vm/sotr/vmhap},\ }\textbf {\bibinfo {volume} {3D-MLSI
  : The program for extraction of 3D inductances of multilayer superconductor
  circuits}} (\bibinfo {year} {2004})}\BibitemShut {NoStop}%
\bibitem [{\citenamefont {Forrester}\ \emph {et~al.}(1994)\citenamefont
  {Forrester}, \citenamefont {Davidson}, \citenamefont {Talvacchio},
  \citenamefont {Gavaler},\ and\ \citenamefont
  {Przybysz}}]{Forrester:1994p2299}%
  \BibitemOpen
  \bibfield  {author} {\bibinfo {author} {\bibfnamefont {M.}~\bibnamefont
  {Forrester}}, \bibinfo {author} {\bibfnamefont {A.}~\bibnamefont {Davidson}},
  \bibinfo {author} {\bibfnamefont {J.}~\bibnamefont {Talvacchio}}, \bibinfo
  {author} {\bibfnamefont {J.}~\bibnamefont {Gavaler}}, \ and\ \bibinfo
  {author} {\bibfnamefont {J.}~\bibnamefont {Przybysz}},\ }\href@noop {}
  {\bibfield  {journal} {\bibinfo  {journal} {Applied Physics Letters},\
  }\textbf {\bibinfo {volume} {65}},\ \bibinfo {pages} {1835} (\bibinfo {year}
  {1994})}\BibitemShut {NoStop}%
\bibitem [{\citenamefont {Bergeal}\ \emph {et~al.}(2008)\citenamefont
  {Bergeal}, \citenamefont {Lesueur}, \citenamefont {Aprili}, \citenamefont
  {Faini}, \citenamefont {Contour},\ and\ \citenamefont
  {Leridon}}]{Bergeal:2008}%
  \BibitemOpen
  \bibfield  {author} {\bibinfo {author} {\bibfnamefont {N.}~\bibnamefont
  {Bergeal}}, \bibinfo {author} {\bibfnamefont {J.}~\bibnamefont {Lesueur}},
  \bibinfo {author} {\bibfnamefont {M.}~\bibnamefont {Aprili}}, \bibinfo
  {author} {\bibfnamefont {G.}~\bibnamefont {Faini}}, \bibinfo {author}
  {\bibfnamefont {J.~P.}\ \bibnamefont {Contour}}, \ and\ \bibinfo {author}
  {\bibfnamefont {B.}~\bibnamefont {Leridon}},\ }\href@noop {} {\bibfield
  {journal} {\bibinfo  {journal} {Nature Physics},\ }\textbf {\bibinfo {volume}
  {4}},\ \bibinfo {pages} {608} (\bibinfo {year} {2008})}\BibitemShut {NoStop}%
\bibitem [{\citenamefont {Gueret}\ \emph {et~al.}(1980)\citenamefont {Gueret},
  \citenamefont {Moser},\ and\ \citenamefont {Wolf}}]{Gueret:1980}%
  \BibitemOpen
  \bibfield  {author} {\bibinfo {author} {\bibfnamefont {P.}~\bibnamefont
  {Gueret}}, \bibinfo {author} {\bibfnamefont {A.}~\bibnamefont {Moser}}, \
  and\ \bibinfo {author} {\bibfnamefont {P.}~\bibnamefont {Wolf}},\ }\href@noop
  {} {\bibfield  {journal} {\bibinfo  {journal} {IBM Journal of Research and
  Development},\ }\textbf {\bibinfo {volume} {24}},\ \bibinfo {pages} {155}
  (\bibinfo {year} {1980})}\BibitemShut {NoStop}%
\bibitem [{\citenamefont {Likharev}\ and\ \citenamefont
  {Semenov}(1991)}]{Semenov:1991}%
  \BibitemOpen
  \bibfield  {author} {\bibinfo {author} {\bibfnamefont {K.~K.}\ \bibnamefont
  {Likharev}}\ and\ \bibinfo {author} {\bibfnamefont {V.~K.}\ \bibnamefont
  {Semenov}},\ }\href@noop {} {\bibfield  {journal} {\bibinfo  {journal} {IEEE
  Trans. Appl. Supercond.},\ }\textbf {\bibinfo {volume} {1}},\ \bibinfo
  {pages} {280} (\bibinfo {year} {1991})}\BibitemShut {NoStop}%
\end{thebibliography}


%

\begin{figure}
\includegraphics[height=15 cm]{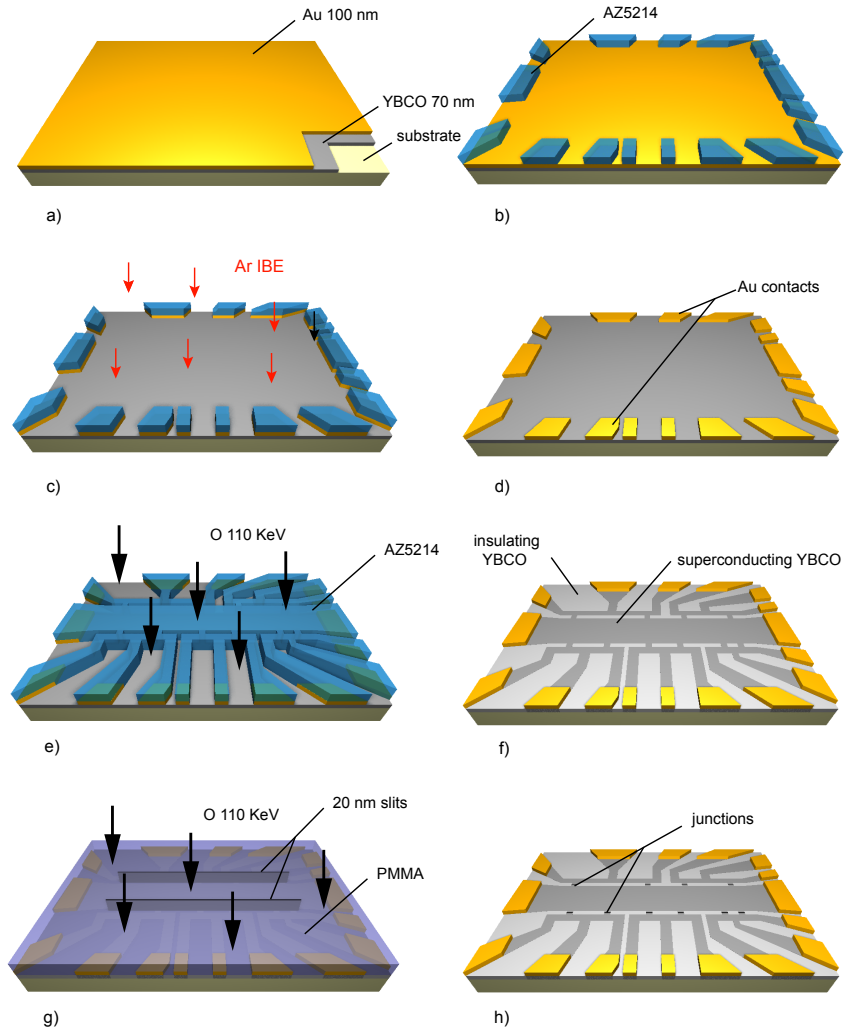}

\caption{Fabrication process. Starting from a 70 nm thick \YBCO (YBCO) film
grown on sapphire covered by an in-situ 100 nm gold layer supplied
by THEVA (a), pads are defined in a AZ5214 negative photoresist (b).
A 500eV Ar Ion Beam Etching (IBE) removes the gold layer leaving only
contact pads (c-d). Squids current and voltage lines are then defined
using again a AZ 5214 photoresist patterning followed this time by
a 110 keV oxygen ion irradiation (e). The use of a high dose of $5\times10^{15}$
ions/cm$^{\text{2}}$ ensures that the surrounding material is changed into an insulator (f). Eventually, josephson junctions are patterned as 20 nm wide slits in a 600 nm
thick poly(methylmethacrylate) (PMMA) photoresist. A subsequent irradiation
with 110 keV oxygen ions and a low dose of $3\times10^{13}$ ions/cm$^{\text{2}}$ gives working
junctions in the temperature range 40-75 K. The electronic beamwritter used in this process is a LEICA EBPG 5000+.}

\end{figure}

\end{document}